\documentclass[useAMS,usenatbib,usegraphicx,letterpaper]{mn2e}

\usepackage{aas_macros}

\usepackage{color}

\newcommand{\myemail}{stephane.corbel@cea.fr}

\def\cx{Cyg~X-3}

\usepackage{lineno}

\defcitealias{Abdo09CX3}{FLC09} 	

\title[The 2011 radio and $\gamma$-ray flare of \cx]{A giant radio flare from Cygnus X-3 with associated  gamma-ray emission}
\author[S. Corbel, G. Dubus et al. ]{S. Corbel$^{1,2}$\thanks{E-mail: \myemail}, G. Dubus$^{3}$, J. A. Tomsick$^{4}$, A. Szostek$^{5,6}$, R. H. D. Corbet$^{7,8}$, 
\newauthor J. C. A. Miller-Jones$^{9}$,  J. L. Richards$^{10}$,  G. Pooley$^{11}$,  S. Trushkin$^{12}$, R. Dubois$^{5}$, 
\newauthor A. B. Hill$^{13}$, M. Kerr$^{5}$, W. Max-Moerbeck$^{10}$,   A. C. S. Readhead$^{10}$,  A. Bodaghee$^{4}$, 
\newauthor V. Tudose$^{14}$, D. Parent$^{15}$,  J.~Wilms$^{16}$, K.~Pottschmidt$^{7,17} $   \\
$^{1}$Universit\'e Paris 7 Denis Diderot and Service d'Astrophysique, UMR AIM,  CEA Saclay, F-91191 Gif sur Yvette, France.  \\
$^{2}$Institut Universitaire de France, 75005 Paris, France.  \\
$^{3}$UJF-Grenoble 1 / CNRS-INSU, Institut de Plan\'etologie et d'Astrophysique de Grenoble (IPAG) UMR 5274, Grenoble, F-38041, France.  \\
$^{4}$Space Sciences Laboratory, 7 Gauss Way, University of California, Berkeley, CA 94720-7450, USA.  \\
$^{5}$Kavli Institute for Particle Astrophysics and Cosmology, Department of Physics and SLAC National Accelerator Laboratory,  Stanford University, Stanford, CA 94305, USA.  \\
$^{6}$Astronomical Observatory, Jagiellonian University, Orla 171, 30-244 Krak\'ow, Poland\\
$^{7}$CRESST and NASA Goddard Space Flight Center, Astrophysics  Science Division, Code 662, Greenbelt, MD 20771, USA \\
$^{8}$Center for Space Science and Technology, University of Maryland  Baltimore County, 1000 Hilltop Circle, Baltimore, MD 21250, USA\\
$^{9}$International Centre for Radio Astronomy Research - Curtin University, GPO Box U1987, Perth, WA 6845, Australia.  \\
$^{10}$Cahill Center for Astronomy and Astrophysics, California Institute of Technology, Pasadena, CA 91125, USA.  \\
$^{11}$Cavendish Laboratory, Cambridge CB3 0HE, UK.  \\
$^{12}$Special Astrophysical Observatory RAS, Karachaevo-Cherkassian Republic, Nizhnij Arkhyz 369167, Russia.  \\
$^{13}$School of Physics \& Astronomy, University of Southampton, Highfield, Southampton,  SO17 1BJ, UK \\
$^{14}$Netherlands Institute for Radio Astronomy, Postbus 2, 7990 AA Dwingeloo, the Netherlands.  \\
$^{15}$Center for Earth Observing and Space Research, College of Science, George Mason University, Fairfax, VA 22030, resident at Naval Research Laboratory, Washington, DC 20375, USA.  \\
$^{16}$Erlangen Centre for Astroparticle Physics, D-91058 Erlangen, Germany.  \\
$^{17}$CRESST and NASA Goddard Space Flight Center, Astrophysics Science Division, Code 661, Greenbelt, MD 20771, USA}

\begin{document}

\date{Accepted for publication in MNRAS }

\pagerange{\pageref{firstpage}--\pageref{lastpage}} \pubyear{2011}

\maketitle

\label{firstpage}

\begin{abstract}
With frequent flaring activity of its relativistic jets, Cygnus X-3 (\cx ) is one of the most active microquasars and is  the only  Galactic black hole candidate with confirmed high energy $\gamma$-ray emission,  thanks to detections by  {\em Fermi}/LAT and {\em AGILE}. 
In  2011, \cx\ was observed to transit to a soft X-ray state, which is  known to be associated with high-energy $\gamma$-ray emission. 
We present the results of a multi-wavelength campaign covering a quenched state, when radio emission from \cx\ is at its weakest and the X-ray spectrum is very soft. A giant ($\sim$ 20 Jy) optically thin radio  flare marks the end of the quenched state, accompanied by rising non-thermal hard X-rays.  {\em Fermi}/LAT observations (E $\ge$100 MeV)  reveal renewed $\gamma$-ray activity associated with this giant radio flare, suggesting a common origin for all non-thermal components.  In addition, current observations unambiguously show that the $\gamma$-ray emission is not exclusively related to the rare giant radio flares. A 3-week period of $\gamma$-ray emission is also detected when \cx\ was  weakly flaring in radio,
 right before transition to the radio quenched state. 
No $\gamma$ rays are observed during the $\sim$ one-month long quenched state, 
 when the radio flux is weakest. 
 Our results suggest transitions into and out of the ultrasoft X-ray (radio quenched) state trigger $\gamma$-ray emission, implying a connection to the accretion process, and also that the $\gamma$-ray activity is related to the level of radio flux (and possibly shock formation), strengthening the connection to the relativistic jets.
\end{abstract}

\section{Introduction}

Galactic and extra-galactic accreting systems containing a neutron star or a black hole {\bf can} produce outflows containing energetic particles that are 
accelerated away from the compact object up to relativistic speeds in collimated jets. These high energy particles, entangled in
the jet magnetic field, lose their energy via synchrotron, inverse Compton  emission  and/or adiabatic losses, or via pion production in the case of baryonic jets, resulting in a broad-band spectrum from radio up to high energy $\gamma$ rays. (e.g. \citealt{Atoyan99,Georganopoulos02,Romero03}).

\cx\ was one of the first sources to be discovered in the early days of X-ray astronomy \citep{Giacconi67}. It is a system consisting of a Wolf-Rayet star  \citep{vK92} and  a compact object (most likely a black hole). With a short 4.8-hr orbital period \citep{Parsignault72}, \cx\  lies at a distance of the order of $\sim$ 7 kpc \citep{Ling09}. The  X-ray spectrum from \cx\ changes between hard and soft states akin to those observed in other accreting X-ray binaries, and is heavily absorbed at low energies by the intervening dense stellar wind \citep{Szostek08a,Hjalmarsdotter09}.  \cx\ is also known for the recurrent activity of its relativistic jets that make it one of the brightest Galactic transient radio sources  (e.g. \citealt{Mioduszewski01,MillerJones04}). 
 
{\it AGILE} and {\em Fermi}/LAT reported concurrent detections of \cx\ in high energy $\gamma$ rays ($>$100 MeV), closely related to the activity of the relativistic jet during the soft X-ray state (\citealt{Tavani09}; \citealt[hereafter FLC09]{Abdo09CX3}).  The $\gamma$-ray emission measured by the {\em Fermi}/LAT was found to be modulated on the orbital period, securing the identification (\citetalias{Abdo09CX3}).  A very short  $\gamma$-ray flare was reported later during a short transient softening of \cx\  in 2010  \citep{Corbel10a, Bulgarelli10a, Williams11b}. No evidence  for emission above 250 GeV has been found by MAGIC \citep{Aleksic10}.

\cx\ is the first binary hosting an accreting compact object and relativistic jet (a.k.a. microquasar) to be detected in $\gamma$ rays. It is clearly an accreting source unlike gamma-ray binaries that are more naturally explained by pulsar spindown power \citep{Dubus06}.  The nature and location of the non-thermal processes that bring particles to high energies, the relationship between these processes, jet launching and accretion state are long-standing questions for both microquasars and active galactic nuclei that stand to benefit from this detection. In \cx, jet emission can be followed and resolved spatially in radio, the accretion state is traced by the X-rays while the newly-detected $\gamma$ rays provide a window into particle acceleration. 
 In early 2011, \cx\ underwent a new transition to the soft state \citep{Kotani11a}, which was accompanied by transient 
$\gamma$-ray emission detected by  {\em AGILE} and  {\em Fermi} \citep{Bulgarelli11a, Bulgarelli11b,Corbel11a}. The transition was monitored in radio, $\gamma$-rays, soft and hard X-rays (\S2). The {\em Fermi}/LAT detections occurred exactly prior to and following a period of quenched radio emission and ultrasoft/hypersoft X-ray emission (\S3). The LAT detection accompanying a major radio flare casts new light on the relationship between non-thermal radio, X-ray and $\gamma$-ray emission and relativistic ejection (\S4). 

\section{Observations and data analysis}

\cx\ is continuously monitored  in $\gamma$-ray, soft and hard X-ray  by all-sky monitors  like {\em Fermi}/LAT, 
{\em MAXI},  {\em RXTE}/ASM and {\em Swift}/BAT. Dedicated radio observations are also frequently conducted 
to constrain the variable activity of its relativistic jets. 

\subsection{$\gamma$-ray}

We  present the results of observations of the Cygnus region  with  the {\it Fermi}  Large Area Telescope (LAT; 
 \citealt{Atwood09a})  along the course of its recent 2011 active phase. The reduction and analysis of LAT data 
were performed using Science Tools v.9.24.  LAT photons within a 15\degr\  acceptance cone centered on the  location of \cx\ 
were selected in the 100 MeV -- 100 GeV energy range. To minimize contamination by Earth albedo photons, $\gamma$-ray events that have reconstructed 
 directions with angles with respect to the local zenith $>$ 100\degr\ have been excluded. The rocking angles were restricted 
 to $<$ 52\degr.  Due to the potential contamination  of  the nearby pulsar PSR~J2032$+$4127 ($\sim$ 30\arcmin\ from \cx), we used 
its most up to date ephemeris\footnote{LAT pulsar timing models available here: https://confluence.slac.stanford.edu/display/GLAMCOG/ LAT+Gamma-ray+Pulsar+Timing+Models} to select the LAT data from its  off-pulse phase intervals (removing only $\sim$ 19 \% of the useful
exposure). The instrument response functions (IRFs)  ``P6\_V11\_DIFFUSE'' have been used throughout this paper.

Aiming to construct a light-curve on $\sim$ daily timescales, we first characterized the sources within this crowded field of view (e.g.
\citetalias{Abdo09CX3}) by using the  {\it Fermi}  data over the two-year period with an internal source list made under 
IRFs ``P6\_V11\_DIFFUSE'' and different spectral models. A binned maximum likelihood  spectral analysis was 
then performed to further constrain the spectral parameters of the sources within 3\degr\  (with all free parameters) and 9\degr\ 
(with only the normalization free) from \cx . 
The two nearby bright pulsars (PSR~J2021$+$4026 and PSR~J2021$+$3651) were modelled with an exponentially cut-off power-law model 
with all parameters let free to vary. We included models for the Galactic  diffuse emission
 (gll\_iem\_v02) and an isotropic component (isotropic\_iem\_v02) including the extragalactic diffuse emission and the residual
background from cosmic rays. Both diffuse components were renormalized for use under IRFs ``P6\_V11\_DIFFUSE''. 

Once the spectral parameters of the sources in the field of view were fully characterized, we fixed those parameters  and performed  an
unbinned maximum likelihood analysis on different timescales ranging from 6 hours to 4 days. In this procedure, the remaining free parameters are 
the normalization of the Galactic diffuse emission and the normalization of  the power-law that is used to model \cx\ (as described in
\citetalias{Abdo09CX3}). In time bins where Cyg X-3 is not detected,  which we take to correspond to a test statistic, TS $<$ 20 (4-day bin)  or TS $<$ 9 (shorter time bins), 
a 95\%  confidence upper limit is calculated.  The upper limit is calculated using the  bayesian method of \citet{Helene91}, as implemented in the Python {\tt UpperLimits} 
module provided with the {\em Fermi} ScienceTools,  with the Cyg X-3 power law photon index fixed to be 2.7.
 
 Following the reactivation of  \cx\ in $\gamma$ rays \citep{Corbel11a}, a dedicated Target of Opportunity (ToO) observation was also conducted 
 by {\it Fermi}/LAT starting on 2011 March 24 (MJD 55644.65). However, the end of the flaring activity of \cx\ resulted in an  termination of the ToO observation
 on March 28 (MJD 55648.63).

\subsection{Radio}

Since the launch of {\it Fermi} in June 2008, we have had an ongoing monitoring program of Galactic binaries with 
the Owens Valley  Radio Observatory (OVRO) 40 m single-dish telescope located in California (USA). 
The OVRO flux densities are measured in a single 3 GHz band centered on 15 GHz using Dicke switching and dual-beam sky switching to remove atmospheric interference. The Galactic binaries sample, including Cyg X-3, is observed with the same cadence and procedures used for the blazar monitoring program described in  \citet{Richards11a}. In addition, two days of intensive monitoring of Cyg X-3 were performed while the source was above the OVRO horizon in February 2011 (MJD 55599 and 55601).  An offset of 
0.124 Jy has been removed from the OVRO flux densities to account for the effect of extended nearby 
sources \citep{Sanchez08a} that are usually resolved out by interferometers (based on  a comparison 
with AMI data obtained simultaneously at the same frequency; \citetalias{Abdo09CX3}).

We take advantage of the large set of measurements provided by the AMI Large  array (Cambridge, UK), consisting of a set of eight 13 m antennas with a 
maximum baseline of $\sim$ 120 m. AMI observations are conducted with a 6 GHz bandwidth  
centered  at 15 GHz \citep{Zwart08}.
We also used the 11.2 GHz data of \cx\ from a monitoring program \citep{Trushkin06} with 
the RATAN-600 telescope located near Zelenchurskaya vil. on the North Caucas (Russia). 
Its 11.2 GHz radiometer  is cooled by the  cryogenic system up to
12 K.
No frequency correction has been
applied to the RATAN data. 

\subsection{X-ray}

We used the one day average quick-look measurements  in soft and hard X-rays from:  1)  the All-Sky 
Monitor aboard the {\em Rossi X-ray Timing Explorer} ({\em RXTE}/ASM; 1.5-12 keV; \citealt{Levine06a}), 2)  the 
{\em Monitor of All-sky X-ray Image} ({\it MAXI}; 1.5-20 keV; \citealt{Matsuoka09a}) installed on the International Space Station, 
and 3) the Burst Alert Telescope on the {\em Swift} mission ({\em Swift}/BAT; 15-50 keV; \citealt{Barthelmy05a}). 

We also obtained pointed {\em RXTE} observations between 2011 February 19 (MJD 55611) and 2011 April 4 (MJD 55655).
The 46 approximately daily monitoring observations performed under this program (P96375) were triggered by Cyg~X-3 entering
the ultra-soft X-ray/quenched radio state, and we followed the evolution of the Cyg~X-3 X-ray properties as the source
made its expected transition out of this state.  We extracted the Proportional Counter Array (PCA; \citealt{Jahoda06})
data from these observations, and the source evolution is characterized in terms of the hardness ratio and the detailed
spectral properties below.  The average PCA exposure time per observation is $\sim$ 2.7 ks.

We produced 3--50~keV PCA energy spectra for the 46 {\em RXTE} observations.  We used
the standard tools provided in HEASOFT v6.11 to produce source and background spectra and included the
recommended 0.5\% systematic uncertainties.  We carried out the spectral fits with the XSPEC v12 software package.
As the spectrum varies with the orbital modulation, we restricted our analysis to the observations obtained
near the peak of the modulation.  Table~1 lists the five observations for which we carried out spectral fits
along with their orbital phases.  The first of these observations occurred when the source was still in the
ultra-soft state and the last was after the gamma-ray flare and during the decay of the radio flare.

We also used the {\em RXTE}/PCA data to characterize the Cyg X-3 
X-ray timing properties during the five observations for which we
reported detailed spectral properties above.  We made light curves
from the ``Standard 1'' data, which has 0.125~s time resolution
and effectively no energy resolution.  The single energy bin 
includes all PCA channels, corresponding to an energy band of 
$\sim$2--60~keV.  We first produced 0.002--4~Hz power spectra
in the \cite{leahy83} normalization and then subtracted the
Poisson noise level and divided by the total count rate to 
convert to the rms power spectra \citep{miyamoto91} that are
shown in Figure~\ref{fig_rms}.  We fitted the power spectra with a 
power-law model, and used these fits to determine the overall
fractional rms level for each observation.

\section{Results}

\begin{figure*}
\includegraphics[angle=0,scale=0.7]{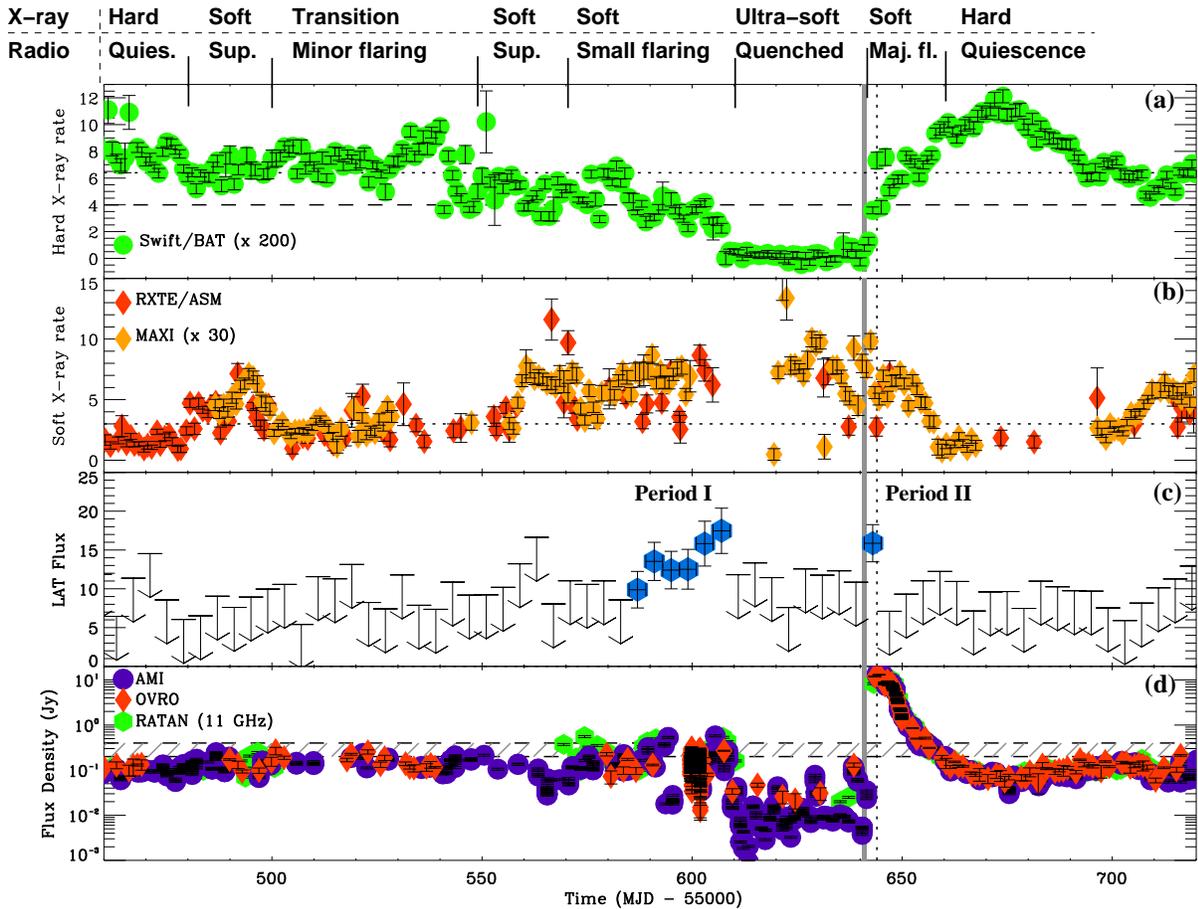}
\caption{Hard X-ray  light curve in panel (a) from {\it Swift}/BAT (15--50 keV; green) with the soft X-ray light curves in panel (b) from {\it RXTE}/ASM (3--5 keV; red)  
and {\em MAXI} (1.5--4 keV; orange)  from 2010 September 21  to  2011 June 8. The horizontal dotted line in (a)  and (b) highlights the emission level corresponding 
to the  hard to soft state transition (e.g.  the ASM 3-5 keV pivotal rate of 3 cts s$^{-1}$ of \citealt{Szostek08a}).  Panel (c):   LAT Flux (E $\ge$  100 MeV) 
light curve of \cx\ obtained in 4-day bins.  The LAT  fluxes  (left axis) are expressed in units of 10$^{-7}$ ph cm$^{-2}$ s$^{-1}$ above 100 MeV. LAT upper limits are 
represented at the 95\% confidence level. 
 Panel (d): Radio lightcurve of \cx\ at 15 GHz based on the OVRO, AMI and RATAN  (at 11.2 GHz)  data. 
The onset of the giant radio flare is indicated in the four panels by the large vertical grey band, whereas the peak of this flare is marked with the vertical dotted line 
(see Figure  \ref{fig_flare}  for a zoom on this interval and Figure \ref{fig_periodI} for period I). 
The  horizontal dashed lines in panels (a) and (d) (and hatched area = the 0.2 to 0.4 Jy zone) highlight  the thresholds (see discussion in paper) corresponding to 
the detection of $\gamma$-ray emission of \cx\
by LAT. The labels in the top of the figure indicate the corresponding X-ray/radio state of \cx\ (``Quies.", ``Sup." and ``Maj. fl." stand respectively for ``Quiescence", 
``Suppressed" and ``Major flaring" radio state). 
\label{fig_lc}}
\end{figure*}

\subsection{Radio and X-ray emission during the 2011 soft state}\label{sect_rad}

The long-term light curves of \cx\ in soft and hard X-rays are presented in the two top panels of Figure \ref{fig_lc},  along with the radio lightcurve in the bottom panel.  
The X-ray and radio states, as defined by  \citet{Szostek08a},  are  marked on top of this Figure.  Up to mid-October 2010 ($\sim$ MJD 55480), \cx\ is found in 
its typical hard state. At that time,  {\em MAXI} and {\em RXTE}/ASM  data show a rise in soft X-rays ($\approx 2-5$ keV) whereas {\em Swift}/BAT measures a decline in 
hard ($>$ 15 keV) X-rays, highlighting an  initial  transition to the soft/suppressed  X-ray/radio state.  The 3-5 keV count rate  from the {\em RXTE}/ASM of $\approx$
  3 cts\,s$^{-1}$ corresponds to the pivotal value separating hard from predominantly soft X-ray states  in \cx\ (see Figure 6 of \citealt{Szostek08a}). 
The decrease ($\sim$ MJD 55500) in soft X-ray flux indicates a temporary return to the level of the hard/soft state transition and minor flaring radio state. 
The further increase in soft X-rays above the pivotal ASM value starting from $\approx$ MJD 55540 marks the full transition to the soft state, while the  abrupt drop in 
hard X-ray flux on MJD 55608 (February 16, 2011) indicates the transition to the ultra-soft X-ray state.

\begin{figure*}
\includegraphics[angle=0,scale=.75]{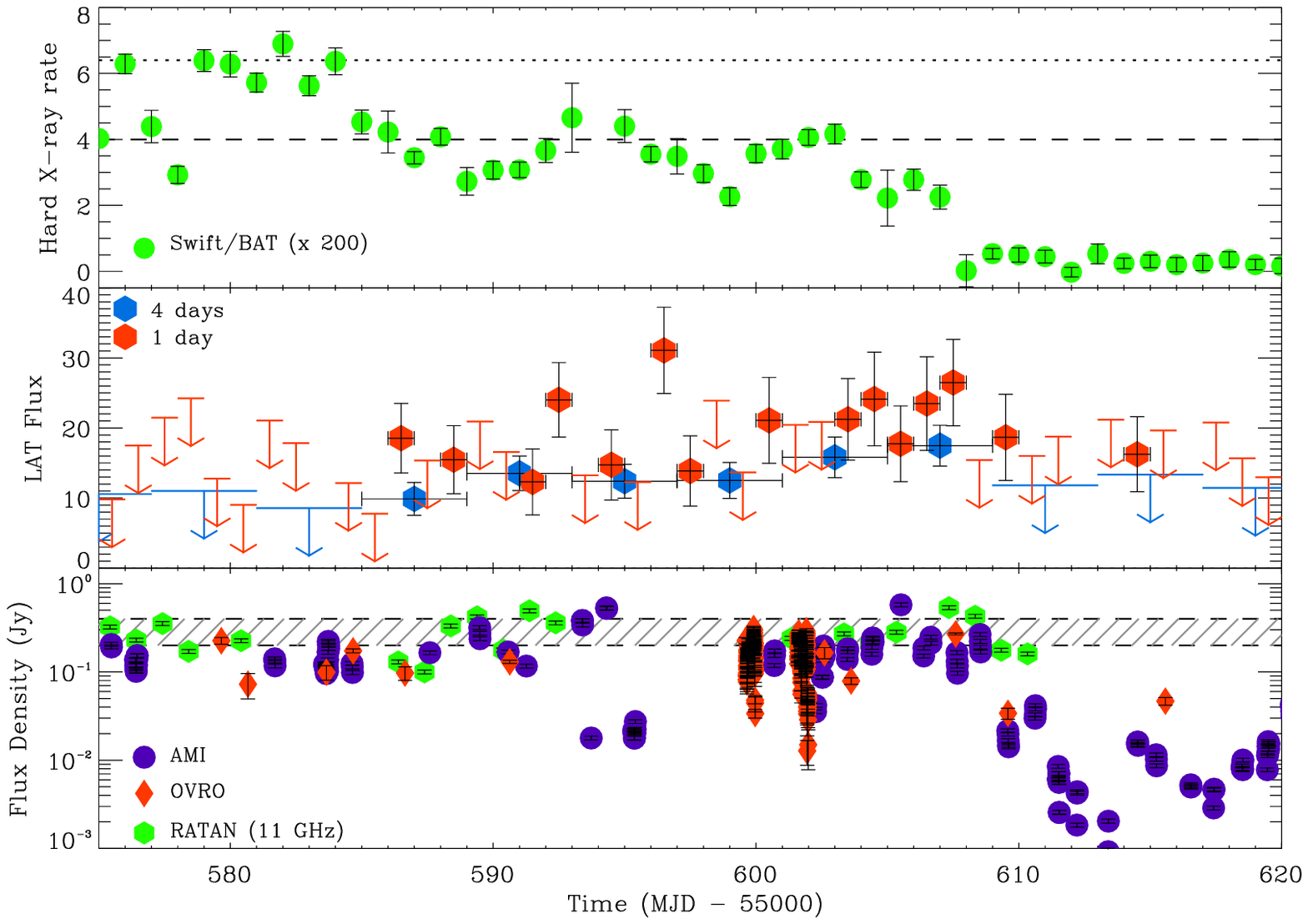}
\caption{Same as Figure 1 but for a time interval around period I (2010 December 30 to 2011 February 28). The soft X-rays (not shown for clarity) were all above the ASM pivotal value of 3 cts $s^{-1}$). The LAT fluxes in  1-day time bin  are also plotted in the middle panel. 
\label{fig_periodI}}
\end{figure*}

 Once fully in the soft state, \cx\ is again in the suppressed radio state  (between  MJD 55540 and 55570), and then  the radio emission is variable with the presence of 
  flares with peak flux densities reaching $\sim$ 0.6 Jy at 15 GHz. 
 As opposed to the solitary major radio flares ($>$ 10 Jy), which are known to follow quenched radio states and to decay on timescales of days, these small radio flares  ($<$ 1Jy)
 are shorter in duration, more rapidly variable and may  occur in clusters of groups that are  alternating with short periods of radio emission at the level typical of  the suppressed radio state
 (a good example can be found in \citealt{MillerJones09b}). 
 The small radio flaring activity was first named by \citet{Waltman94} and was not included in the classification of X-ray and radio states of \citet{Szostek08a}.
 Apparent superluminal expansion during two small flares, with peak flux up to  0.3 Jy at 15 GHz, was observed by  \citet{Newell98}.  Furthermore, on several occasions
 these small flares have been observed up to the millimeter domain,  which was also occasionally the case early in 2011 with inverted spectra up to 98 GHz (e.g. \citealt{Kotani11b}; 
 S. Trushkin private com.), suggesting temporary optically thick synchrotron emission. 
 All these radio properties are consistent with \cx\ moving in early 2011 from  the suppressed state to a state of small flaring radio activity.

The time span between MJD 55610 and  55640 with weak radio emission (down to 2 mJy) and very low hard X-ray emission is characteristic of the quenched radio state that always precedes a giant flare \citep{Waltman94, Fender97a}. \cx\ is then in its ultrasoft \citep{Szostek08a,Koljonen10} X-ray spectral state characterized by strong thermal soft X-rays and a very weak or absent non-thermal power-law in hard X-rays.
 Optically thick radio spectra are usually observed in this state \citep{Waltman95}.

The giant radio flare that ends the quenched state reached a flux of almost 20 Jy  at 15 GHz \citep{Corbel11a}.  Figure \ref{fig_flare} highlights the rising part of this flare, indicating a likely onset (defined by the gradual increase of radio emission at the end of the quenched radio state) around MJD 55641.0 $\pm$ 0.5 and peak flux around MJD 55644 (also marked on 
Figure \ref{fig_lc}). The delay between onset and peak is consistent with observations  during previous giant radio flares (of the order of 2 to 4 days, e.g. \citealt{Waltman95, MillerJones04}).  We caution that the exact trigger time of this relativistic ejection event could possibly occur after, or even before, the radio onset we highlighted above.  Alternatively, the detection of a non thermal X-ray component by {\it RXTE}/PCA on  MJD 55642.0 (Fig.~4, Table 1  and  section 4.2)  and the hardening of the X-ray spectra (Fig.~3) may  also signal the trigger time of the ejection event (indicated on Figure \ref{fig_flare}). 
The  precise ejection date should be better constrained with the VLBI campaign we conducted during this flare (Miller-Jones et al. in prep.). 

The hard X-rays switch back on and are correlated with the radio flux during major flares \citep{Szostek08a}. The {\em RXTE}/PCA hardness ratio (Figure \ref{fig_flare}) confirms the hardening, as well as the  appearance of a  hard X-ray tail in the {\em RXTE}/PCA spectra  (see also Figure~\ref{fig_pca} and discussion in section 4.2) during the onset of the flare. 
This hard X-ray tail was not present during the radio quenched period of the 2011 soft state (Figs.~\ref{fig_lc} and \ref{fig_pca}). 
The spectra during major radio flares typically show a non-thermal power-law in hard X-rays  \citep{Szostek08a,Koljonen10} and optically thin radio emission \citep{Waltman95}.

\subsection{Flaring $\gamma$-ray emission from \cx }

The {\em Fermi}/LAT $\gamma$-ray light curve (0.1-100 GeV) with 4-day time bins is presented in panel (c) of  Figure \ref{fig_lc}.  Only the very significant data-points  with test statistic \citep{Mattox96} TS $\ge$ 20 ($\sim$ 4.5 $\sigma$) are plotted, otherwise, they are represented  as upper limits.  Two distinct phases of $\gamma$-ray activity are found: (1) a long phase  ($\sim$ 3 weeks) of  $\gamma$-ray activity just before the ultra-soft state (hereafter period I); (2) a shorter $\gamma$-ray active phase  ($\leq 5$ days)  in conjunction with the giant radio flare (hereafter period II). No $\gamma$-ray emission is detected during the quenched radio state in between period I and II,  corresponding to the ultra-soft  X-ray state.

Regarding period I (see also zoom in Fig. \ref{fig_periodI}), we note that the 4-day bin data  (from  MJD 55585 to 55610) are consistent with a steady increase of $\gamma$-ray emission 
up to a flux ($\ge$100 MeV)  $\sim$ 1.8  $\times$ 10$^{-6}$ ph cm$^{-2}$ s$^{-1}$,  just before transition to the quenched radio state. However, the data on shorter bins (Fig. \ref{fig_periodI}) 
indicate variability  on timescales as short as one day. 
$\gamma$-ray emission in period I  occurs after the soft X-ray emission has increased beyond the ASM pivotal value ($\sim$ 3 cts s$^{-1}$),  also during a significant decrease in hard X rays (BAT flux $<$ $\sim$ 0.02 cts  cm$^{-2}$ s$^{-1}$) and in a period of short and faint radio flaring, placing \cx\ in the small flaring radio state  (as discussed in section \ref{sect_rad}). The {\it AGILE} collaboration also reported detections of \cx\  during  period I \citep{Bulgarelli11a, Bulgarelli11b}. No pointed {\it RXTE} observations were executed in period I, and therefore we cannot determine if a non-thermal hard X-ray tail was present during that time, as we do below for period II.

\begin{figure*}
\includegraphics[angle=0,scale=.75]{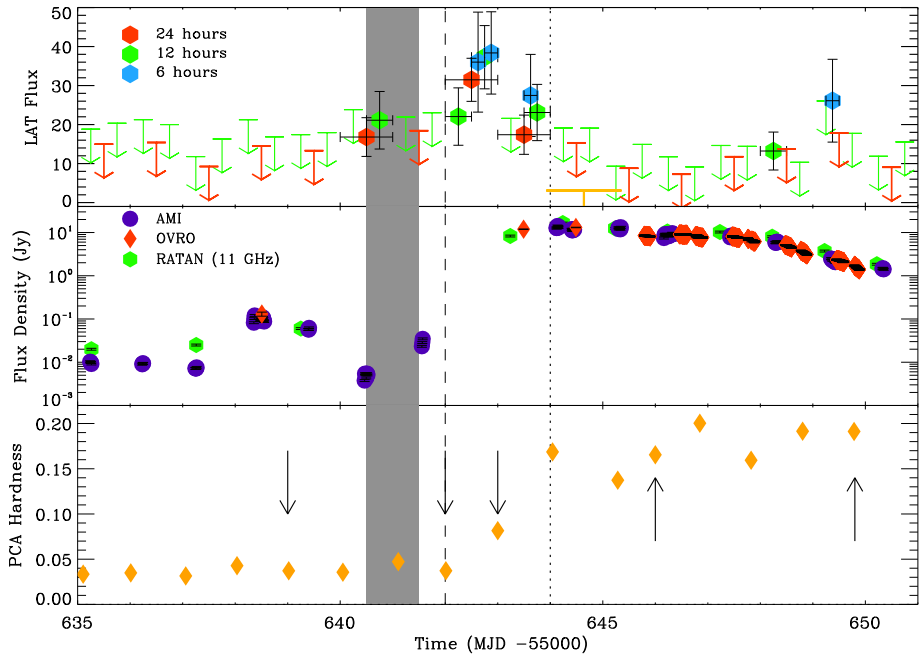}
\caption{Top Panel: LAT  flux (E $\ge$ 100 MeV) light curve of \cx\ obtained in 1-day, 12-hour, and 6-hour bins for the  period around the giant radio flare
(2011 March 15 to 31). The  orange upper limit corresponds to the 95\% confidence upper limit during the {\it Fermi} ToO observations.
Middle Panel: Radio lightcurve (same data as Figure 1, including AMI, OVRO and RATAN) of \cx\ highlighting the  onset of the giant radio flare,  that gives a starting 
date  for the radio flare of MJD $\sim$ 55641.0 $\pm$0.5. 
Bottom Panel: Evolution of the {\it RXTE}/PCA hardness ratio (defined as count rate in band 10-50 keV over  band 3-10 keV). The error bars are smaller 
than the symbol size. The arrows indicate the time of the five {\it RXTE} spectra discussed in section 4.2 and  Fig.~\ref{fig_pca}. For all panels, the onset of the 
giant radio flare is indicated by the large vertical grey band (based on the increase in radio flux density) or the vertical dashed line (based on the appearance 
of a non thermal X-ray component), whereas the peak of this flare is marked with 
the vertical dotted line.
\label{fig_flare}}
\end{figure*}

For period II, we constructed  LAT light curves on 6-hour, 12-hour and 1-day intervals to highlight the onset of the $\gamma$-ray emission and to allow a comparison with the giant radio flare. Due to the shorter integration time, only data-points with TS $>$ 9  ($\sim$ 3 $\sigma$) are plotted (otherwise, they are represented as upper limits)  in Figure \ref{fig_flare}. The short $\gamma$-ray activity in period II occurs during the rise of the radio flare. The $\gamma$-ray emission turns off before the peak of radio emission (vertical dotted line in  Figure \ref{fig_flare}). 
The first LAT detection occurs on MJD 55640.5 $\pm$ 0.5, implying  a $\gamma$-ray turn-on  consistent  with the exact onset of the radio flare  defined by the increase in radio 
flux density (illustrated by the grey area in Figure \ref{fig_flare}; see also section 3.1). 
Based on the 6-hour and 12-hour bin lightcurves, the peak in $\gamma$ rays ($\approx$ 3 to 4  $\times$ 10$^{-6}$ ph cm$^{-2}$ s$^{-1}$) is reached on MJD 55642.75 $\pm$ 0.25, which is well before (by 1.5 day) the peak flux in radio. Short timescale variability might be present within this active phase: the LAT 1-day bin on MJD 55641.5 $\pm$ 0.5 is consistent with a  non-detection although there is no significant change in LAT exposure towards \cx\ on this specific day. This would be consistent with the upper limits reported by {\it AGILE}  around this time \citep{Bulgarelli11c}.  Furthermore, we note that most of the LAT 
detections occur after the onset of the non-thermal X-ray component (illustrated by the dashed line in Figure \ref{fig_flare}; see also section 3.1). 
{\it Fermi} conducted a dedicated ToO observation from MJD 55644.65 to 55648.63 with no detection of \cx. We derived a 95\% LAT flux ($\ge$100 MeV)  upper limit of 3.1 $\times$ 10$^{-7}$ ph cm$^{-2}$ s$^{-1}$ for the ToO interval, implying a reduction
of the $\gamma$-ray emission by more than a factor 10 on a timescale of a few days. Furthermore, we find no significant detection by the LAT at the time of the other detection reported by 
 {\it AGILE} \citep{Piano11a} on MJD 55710.

 Integrating  the LAT data separately over the two flaring periods results in photon indices $\sim$ 2.5--2.7 consistent with  the previously published LAT spectra \citepalias{Abdo09CX3}. 
Furthermore, modulation of the gamma-ray emission at the orbital period of \cx\ \citepalias{Abdo09CX3} is again detected  by {\it Fermi}/LAT in both epochs; but the  short 
duration of the present activity  prevents a detailed analysis and will be investigated in a subsequent study using Pass 7 
IRFs\footnote{Information available here: http://fermi.gsfc.nasa.gov/ssc/ data/analysis/documentation/Pass7\_usage.html} 
which have
increased effective area at low energies.

 \section{Discussion} \label{sect_discus}

\begin{table*}
\begin{minipage}{126mm}
\begin{tabular}{cccccc}
\multicolumn{6}{c}{Spectral and timing parameters for the 5 RXTE observations of \cx }\\
\hline \hline
Parameter & Spectrum 1 & Spectrum 2 & Spectrum 3 & Spectrum 4 & Spectrum 5\\ 
\hline 
ObsID & 29-00 & 32-00 & 33-00 & 36-00 & 40-00\\
MJD-55000 & 639.0 & 642.0 & 643.0 & 646.0 & 649.8\\
Orbital Phase & 0.46-0.65 & 0.51-0.64 & 0.41-0.60 & 0.46-0.65 & 0.43-0.60\\
\hline  \hline
Spectral  & & & & & \\ \hline 

$l_{\rm h}/l_{\rm s}$ & $0.15^{+0.05}_{-0.01}$ & $0.28^{+0.08}_{-0.05}$ & $0.018^{+0.008}_{-0.005}$ & $0.061^{+0.007}_{-0.009}$ & $0.034\pm 0.016$\\
$kT_{\rm bb}$ (eV) & $319^{+254}_{-91}$ & $291^{+189}_{-91}$ & $247^{+12}_{-22}$ & $285^{+18}_{-41}$ & $257^{+15}_{-35}$\\
$l_{\rm nt}/l_{\rm h}$ & $<$$9\times 10^{-4}$ & $0.013^{+0.002}_{-0.003}$ & $0.22^{+0.01}_{-0.03}$ & $0.44^{+0.26}_{-0.12}$ & $0.996^{+0.004}_{-0.500}$\\
$\tau_{\rm p}$ & $3.6^{+1.0}_{-0.1}$ & $7.7^{+1.7}_{-1.0}$ & $0.026^{+0.008}_{-0.001}$ & $0.049^{+0.009}_{-0.019}$ & $0.030^{+0.069}_{-0.023}$\\
$\Gamma_{\rm inj}$ & $4.5^{+0.5}_{-4.5}$ & $4.0^{+0.0}_{-1.1}$ & $5.0^{+0.0}_{-0.5}$ & $5.8^{+0.2}_{-0.3}$ & $5.8^{+0.2}_{-0.2}$\\
$\chi^{2}/\nu$ & 78/62 & 57/62 & 52/63 & 61/62 & 46/62\\
\hline \hline
\bf
Timing & & & & & \\  \hline
$\alpha$ & $1.97\pm 0.20$  & $3.2^{+2.1}_{-0.9}$ &  $2.7\pm 0.4$ & $2.29^{+0.13}_{-0.14}$& $2.14^{+0.24}_{-0.27}$ \\  
r.m.s. (\%) & $2.21^{+0.02}_{-0.14}$& $0.75^{+0.01}_{-0.23}$ & $1.73^{+0.02}_{-0.18}$ & $3.73^{+0.13}_{-0.29}$ & $2.09^{+0.12}_{-0.30}$\\  

\hline 

\end{tabular}
\caption{Main spectral parameters (according to their definitions in {\tt eqpair}; see \citealt{Hjalmarsdotter09}) from  the best fitting models of the five {\it RXTE} spectra discussed in section 4 and Fig.~\ref{fig_pca}. 
The  {\tt eqpair} parameters: $l_{\rm h}/l_{\rm s}$ defines the ratio of the hard to soft compactness, $kT_{\rm bb}$ is the temperature of the inner edge of the accretion disk, 
$l_{\rm nt}/l_{\rm h}$ corresponds to the  fraction of power supplied to the energetic particles which goes into accelerating non-thermal particles, $\tau_{\rm p}$ is the 
thermal  plasma optical depth and  $\Gamma_{\rm inj}$ is the index of the  power-law of the accelerated non thermal electrons.  For the timing section in the bottom of the table, 
we report  the power-law index $\alpha$ used in the fitting of the power density spectra (see Figure \ref{fig_rms}) along with the  rms amplitude (in \%) of the variability in the 0.002-0.1 Hz range.  The quoted uncertainties correspond  to 90\% confidence.}
\end{minipage}
\end{table*}

\subsection{Conditions for detection of high energy $\gamma$-ray emission from \cx }

High-energy $\gamma$-ray emission has now been reported by {\em Fermi}/LAT at four different epochs (Oct./Dec. 2008, Jun./Jul. 2009, May 2010, March 2011). In all cases the LAT detections were contemporaneous, but not coincident, with the hard X-ray lightcurve crossing a threshold level,   {\em i.e.} {\em Swift}/BAT count rates of 0.02 cts  cm$^{-2}$ s$^{-1}$ (if only for a day in 2010, \citealt{Williams11b}). In period I, the crossing proceeded from higher to lower flux, whereas in period II it was in the opposite direction. 
The {\em RXTE}/ASM flux in the 3-5 keV band was also always at or above the pivotal value of 3\,cts\,s$^{-1}$ that separates the canonical hard and soft X-ray states of \cx\ \citep{Szostek08a,Koljonen10}. These conditions on the X-ray fluxes appear necessary to detect $\gamma$-ray emission (see also \citealt{Piano11b} for the AGILE detections). 

When its X-ray spectrum is predominantly soft, \cx\ can be found in a suppressed, quenched, small or major flaring state depending on the behavior in radio. The 2008-9 detections and the new LAT detection in period II occurred during major flaring activity, with radio fluxes in excess of a few Jy and optically thin spectra. The new LAT detection in period I occurred at a time when \cx\ was most likely moving through a series of small radio flares, with 15 GHz  flux densities  $\approx 0.1$-0.6 Jy. The radio and X-ray conditions during the brief $\gamma$-ray detection of May 2010 \citep{Williams11b} appear similar to those of period I. We do not detect  $\gamma$-ray emission during the quenched radio state,  when the radio emission is very low and the X-ray spectrum is ultrasoft. However, we can not exclude that very faint $\gamma$-ray emission (at least much below the sensitivity of LAT) could eventually be present in this state, 
as the radio lightcurve (Fig.~1) indicate some variability from the jets.  We also note that $\gamma$ rays are not present in the mid to later stages of period II, even though the radio 
emission is still very bright ($>$ 1 Jy). Slowly decaying jets are not associated with $\gamma$-ray emission, and consequently a rapid rise in radio emission (possibly caused by strong shocks, see section \ref{sect_shock}), up to at least   $\approx$ 0.2 to 0.4  Jy, may therefore be required. 

Hence, the level of $\gamma$-ray emission depends also on the presence of radio emission at a level greater than $\approx$ 0.2 to 0.4  Jy, reinforcing the association of $\gamma$-ray activity with the significant emission from the relativistic jets. 
In other words, particle acceleration to very high energies does not happen exclusively during rare major radio flares but a relatively low level radio activity suffices.
To summarize, three conditions seem to be required in order to detect significant high energy $\gamma$-ray emission from \cx : (1) a high level of soft X-ray emission (the 3-5 keV  {\it RXTE}/ASM value above 3\,cts\,s$^{-1}$, i.e. \cx\ needs to be in the soft state), (2) a low level of hard X-ray emission ({\em Swift}/BAT below 0.02 cts  cm$^{-2}$ s$^{-1}$),  and (3) the presence of significant emission  with rapid variation from active  relativistic jets (with 15 GHz radio flux above $\approx$ 0.2 to 0.4  Jy).  However,  $\gamma$-ray emission can temporarily be seen at lower radio fluxes (e.g. onset of emission during 
epoch II), and therefore a comprehensive picture taking into account possible delays or binning will be examined in future work.  Furthermore, it is unclear at present how the shape of the radio spectrum (optically thin or thick synchrotron emission) relates to the $\gamma$-ray emission. 

\subsection{A non thermal X-ray component during the giant radio flare}

\begin{figure}
\includegraphics[angle=0,scale=.45]{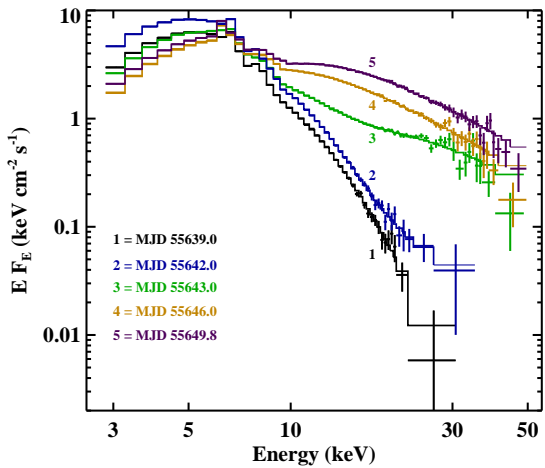}
\caption{ {\it RXTE}/PCA spectra during the soft to hard state transition associated with the giant radio flare of \cx\ in 2011. They correspond to the periods:  before the 
onset of the radio and $\gamma$-ray flare (spectrum 1 on MJD 55639),  (2) the onset of the  $\gamma$-ray flare (spectrum 2 on MJD 55642), (3) the peak of the
 $\gamma$-ray flare (spectrum 3 on MJD 55643), and (4) after the  peak of the radio flare (spectrum 4 and 5).  For illustrative purposes, adjacent bins have been grouped 
 until they reach a significant detection of at least 3$\sigma$, but  with a maximum of 10 bins combined. 
 \label{fig_pca}}
\end{figure}

Giant radio flares start in the ultrasoft X-ray state, which also shows a rising non-thermal component beyond 15 keV while the soft thermal emission remains steady (e.g. \citealt{Szostek08a}). The {\em RXTE}/PCA spectra during the transition (Figure~\ref{fig_pca}) confirm the appearance of a  hard X-ray tail during the onset of the flare in period II. The March 2011 giant radio flare shows that the simultaneous rise in non-thermal radio and hard X-ray components also involves high-energy $\gamma$-ray emission. 
One of the {\em RXTE}/PCA  observations occurred on MJD 55643 (spectrum 3 in Figure 4 and Table 1) while the source was being detected in the $\gamma$-ray band by LAT.
Due to the complicated shape of the X-ray spectrum, the extrapolation from the X-ray to the $\gamma$-ray band is not straightforward.  However,  a rough estimate can be made 
by re-fitting the 10-50~keV portion of the PCA spectrum with a power-law. The value of the best fit photon index is $\Gamma = 3.4$, and an extrapolation to $>$100~MeV 
predicts a flux of $1.0\times 10^{-11}$ ph~cm$^{-2}$~s$^{-1}$.  This is several orders of magnitude lower than the measured LAT flux, indicating that the X rays and $\gamma$ rays cannot be
part of the same power-law (even if the two emission regimes could still be connected) and that a spectral break is located between  $\sim$ 100 keV and 100 MeV (see also \citealt{Zdziarski11}). If the hard X rays and $\gamma$ rays are due to the same radiative process and generated at the same location then the orbital modulation at both frequencies should be in phase.

\begin{figure}
\includegraphics[angle=0,scale=.45]{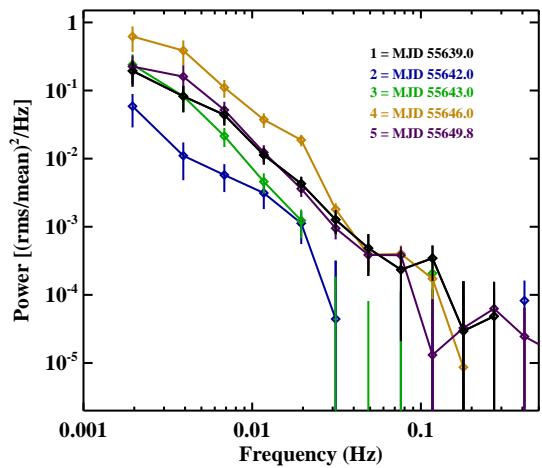}
\caption{   {\it RXTE}/PCA power density spectra during the soft to hard state transition (same dates and colors as in Figure \ref{fig_pca}). 
 \label{fig_rms}}
\end{figure}

It is well known that \cx\ has a highly complex X-ray spectrum, and, as expected, simple models such as a
power-law, a power-law with a cutoff, or thermal Comptonization (all with simple absorption) provide 
very poor fits.  Thus, we used a model that has previously been used for \cx\  \citep{Szostek08b,Hjalmarsdotter09},
which is based on the {\tt eqpair} hybrid thermal/non-thermal Comptonization model \citep{Coppi92,Coppi99}.  In addition
to {\tt eqpair}, we included absorption with partial covering with two separate values of $N_{\rm H}$ and iron features
(a broad line, a narrow line, and an edge) as described in \citet{Hjalmarsdotter09}.  As shown by the $\chi^{2}$
values in Table 1, this model provided acceptable fits.

The  {\tt eqpair } parameters from the fits to the X-ray spectra are given in Table 1. Whereas the  purpose of this work is not to do 
detailed spectral modeling (e.g. see \citealt{Szostek08b,Hjalmarsdotter09}),  this information can  be used to provide some insights into the evolution  of the X-ray spectra during the major radio flare (period II). However, as we do not have X-ray data below 3 keV, several degeneracies within the  {\tt eqpair }  model (e.g. the hydrogen column density and the inner 
accretion disk temperature) cannot be removed. Whereas the model provides good fits to the data, the values reported in Table 1 should not be over-interpreted. 
 The solutions we obtained are in the regime of low seed photon temperature $kT_{\rm bb}$, very soft electron injection spectrum $\Gamma_{\rm inj}$ and 
 low plasma optical depth $\tau_{\rm p}$.   Within this regime, one trend that is notable is the gradual increase  in the $l_{\rm nt}/l_{\rm h}$ parameter.
This appears to be an indication of the increasing importance of non-thermal Comptonization in the spectrum as \cx\ leaves the ultra-soft state.  It is not clear how this is related to the gamma-ray emission since the  LAT detected \cx\ only during the acquisition of spectrum 3, and this is not the spectrum with the largest non-thermal X-ray contribution (unless delayed emission
is involved). 

The  power density spectra (PDS) corresponding to the five {\it RXTE}/PCA observations discussed above are plotted in Fig. \ref{fig_rms}. The PDS at frequencies
higher than 10$^{-3}$ Hz are well described by a power-law with an index of $\sim$ --2 (see the timing  parameters in Table 1). No signal is detected above 0.1 Hz. 
The slope and rms  noise levels of the PDS are similar to previous studies \citep{Choudhury04,Axelsson09}.  One noticeable difference can be seen in the power spectrum \#2 with 
a  lower  noise level. This observation corresponds to the onset of non-thermal Comptonization as discussed in the previous paragraph. This lower variability may possibly be associated with a quiet accretion disk
between the onset of the radio flare and the formation of the corona producing the hard X-ray emission.  

\subsection{Connecting the relativistic jets  and the high energy emission in \cx }\label{sect_shock}

A coherent picture of the link between accretion, ejection and the non-thermal emission we observe has yet to emerge. High levels of soft X-rays together with significant radio emission are related to the $\gamma$-ray activity but the relative timings at the various wavelengths remain confusing: particle acceleration and cooling may lead to differing lags and peaks; correlations may also be blurred when multiple flares superpose. Rising soft X-ray emission is thought to accompany an increased accretion rate (perhaps due to variations in mass loss rate from the Wolf-Rayet companion star, e.g. \citealt{Gies03}) and a decreasing inner disk radius, eventually quenching the jet in the ultrasoft state (e.g. \citealt{Hjalmarsdotter09}). 

VLBI observations have shown that the variations in radio flux during major flares come from the resolved jet on milliarcsecond ($\geq$ a.u.) scales and not from the core \citep{Tudose10}, raising the intriguing possibility that the non-thermal emission region is outside the binary system during major flares. However, there is no evidence yet that the radio emission is detached from the core during the {\em rising} phase of a major flare (when gamma-ray emission was detected). Locating the high energy electrons very far away (e.g. via  VLBI observations) 
would place stringent constraints on jet parameters if the observed $\gamma$-ray modulation is due to inverse Compton upscattering of photons from the Wolf-Rayet star \citep{Dubus10,Sitarek11,Zdziarski11}. 

A possible scenario is that the non-thermal emission is related to shocks forming at various distances along the jet, as previously suggested by modeling of the radio activity \citep{Lindfors07,MillerJones09b}.  Transitions in/out of the ultrasoft X-ray state then signal a decrease/increase in jet efficiency with the non-thermal region moving in/out. Gamma-ray emission may be most efficient at some ``sweet-spot'' distance bounded by strong pair production on thermal X-rays \citep{Cerutti11,Sitarek11} and a declining seed photon density for inverse Compton scattering \citep{Dubus10,Zdziarski11}. Detections prior to and after the quenched state would be due to the shock moving through this region as the jet turns off/on.

The application of the shock-in-jet model to \cx\  strongly suggests that the faster, weaker 
radio flares (like in period I) occur closer to the core, whereas the brighter radio flares (like in period II) occur further downstream \citep{MillerJones09b}. A  shock closer to the core during period I than during period II is also consistent with the brighter $\gamma$-ray emission that is observed in period I, assuming a jet with constant speed. The energy density in seed photons decreases if shocks occur far downstream, reducing the inverse Compton luminosity in period II. An additional signature should be a stronger $\gamma$-ray modulation when shocks occur close to the core, hence close to the Wolf-Rayet companion star.

\section*{Acknowledgments}
We thank the {\it Fermi}  team for accepting and promptly conducting the LAT Target of Opportunity on \cx\ in March 2011. We acknowledge  Elmar Koerding, Andrzej Zdziarski,  and the 
referee for their comments on the manuscript.  The research by SC leading to these results has received funding from the European Community (EC) Seventh Framework Programme (FP7/2007-2013) under grant agreement number ITN 215212 ÓBlack Hole Universe.  GD acknowledges support from the EC via contract ERC-StG-200911. JAT acknowledges partial support from NASA {\it Fermi}  Guest Observer award
NNX10AP83G and from NASA Astrophysics Data Analysis Program award NNX11AF84G. 
The {\it Fermi} LAT Collaboration acknowledges support from a number of agencies and institutes for both development and the operation of the LAT as well as scientific data analysis. These include NASA and DOE in the United States, CEA/Irfu and IN2P3/CNRS in France, ASI and INFN in Italy, MEXT, KEK, and JAXA in Japan, and the K.~A.~Wallenberg Foundation, the Swedish Research Council and the National Space Board in Sweden. Additional support from INAF in Italy and CNES in France for science analysis during the operations phase is also gratefully acknowledged.   This research has made use of the MAXI data provided by RIKEN, JAXA and the MAXI team. Swift/BAT transient monitor results provided by the Swift/BAT team. We also acknowledge the RXTE/ASM team for the X-ray monitoring ASM data. AMI is supported by STFC and the University of Cambridge. The OVRO 40~m monitoring program was supported in part by NASA grants NNX08AW31G and NNG06GG1G and NSF grant AST-0808050. The RATAN-600 observations were carried out with the financial support of
the Ministry of Education and Science of the Russian Federation.

\bibliographystyle{mn2e_fixed}

\label{lastpage}

\end{document}